\documentclass[aps, prl, reprint, superscriptaddress,showpacs,amsmath,amssymb, twocolumn]{revtex4}

\usepackage{graphicx}

\begin{document}
\title{Doping-dependent nodal Fermi velocity in Bi-2212 revealed by high-resolution ARPES}
\author{I. M. Vishik}
\affiliation {Stanford Institute for Materials and Energy Sciences, SLAC National Accelerator Laboratory, 2575 Sand Hill Road, Menlo Park, CA 94025, USA}
\affiliation{Geballe Laboratory for Advanced Materials, Departments of Physics and Applied Physics, Stanford University, Stanford, CA 94305, USA}
\author{W. S. Lee}
\affiliation {Stanford Institute for Materials and Energy Sciences, SLAC National Accelerator Laboratory, 2575 Sand Hill Road, Menlo Park, CA 94025, USA}
\affiliation{Geballe Laboratory for Advanced Materials, Departments of Physics and Applied Physics, Stanford University, Stanford, CA 94305, USA}
\author{F. Schmitt}
\affiliation {Stanford Institute for Materials and Energy Sciences, SLAC National Accelerator Laboratory, 2575 Sand Hill Road, Menlo Park, CA 94025, USA}
\affiliation{Geballe Laboratory for Advanced Materials, Departments of Physics and Applied Physics, Stanford University, Stanford, CA 94305, USA}
\author{B. Moritz}
\affiliation {Stanford Institute for Materials and Energy Sciences, SLAC National Accelerator Laboratory, 2575 Sand Hill Road, Menlo Park, CA 94025, USA}
\affiliation{Geballe Laboratory for Advanced Materials, Departments of Physics and Applied Physics, Stanford University, Stanford, CA 94305, USA}
\author {T. Sasagawa}
\affiliation {Materials and Structures Laboratory, Tokyo Institute of Technology, Meguro-ku, Tokyo 152-8550, Japan}
\author{S. Uchida}
\affiliation {Department of Physics, Graduate School of Science, University of Tokyo, Bunkyo-ku, Tokyo 113-0033, Japan}
\author{K. Fujita}
\affiliation {Laboratory for Atomic and Solid State Physics, Department of Physics, Cornell University, Ithaca, NY 14853, USA}
\author{S. Ishida}
\affiliation {Department of Physics, Graduate School of Science, University of Tokyo, Bunkyo-ku, Tokyo 113-0033, Japan}
\author{C. Zhang}
\affiliation{State Key Laboratory of Crystal Materials, Shandong University, Jinan, 250100, P.R.China}
\author{T. P. Devereaux}
\affiliation {Stanford Institute for Materials and Energy Sciences, SLAC National Accelerator Laboratory, 2575 Sand Hill Road, Menlo Park, CA 94025, USA}
\affiliation{Geballe Laboratory for Advanced Materials, Departments of Physics and Applied Physics, Stanford University, Stanford, CA 94305, USA}
\author{Z. X. Shen}
\affiliation {Stanford Institute for Materials and Energy Sciences, SLAC National Accelerator Laboratory, 2575 Sand Hill Road, Menlo Park, CA 94025, USA}
\affiliation{Geballe Laboratory for Advanced Materials, Departments of Physics and Applied Physics, Stanford University, Stanford, CA 94305, USA}


\date{\today}

\begin{abstract}
The improved resolution of laser-based angle-resolved photoemission spectroscopy (ARPES) allows reliable access to fine structures in the spectrum. We present a systematic, doping-dependent study of a recently discovered low-energy kink  in the nodal dispersion of Bi$_2$Sr$_2$CaCu$_2$O$_{8+\delta}$  (Bi-2212), which demonstrates the ubiquity and robustness of this kink in underdoped Bi-2212.  The renormalization of the nodal velocity  due to this kink becomes stronger with underdoping, revealing that the nodal Fermi velocity is non-universal, in contrast to assumed phenomenology.  This is used together with laser-ARPES measurements of the gap velocity, v$_2$, to resolve discrepancies with thermal conductivity measurements.
\end{abstract}

\maketitle

\begin{figure} [t]
\includegraphics [type=jpg,ext=.jpg,read=.jpg,clip, width=3.3 in]{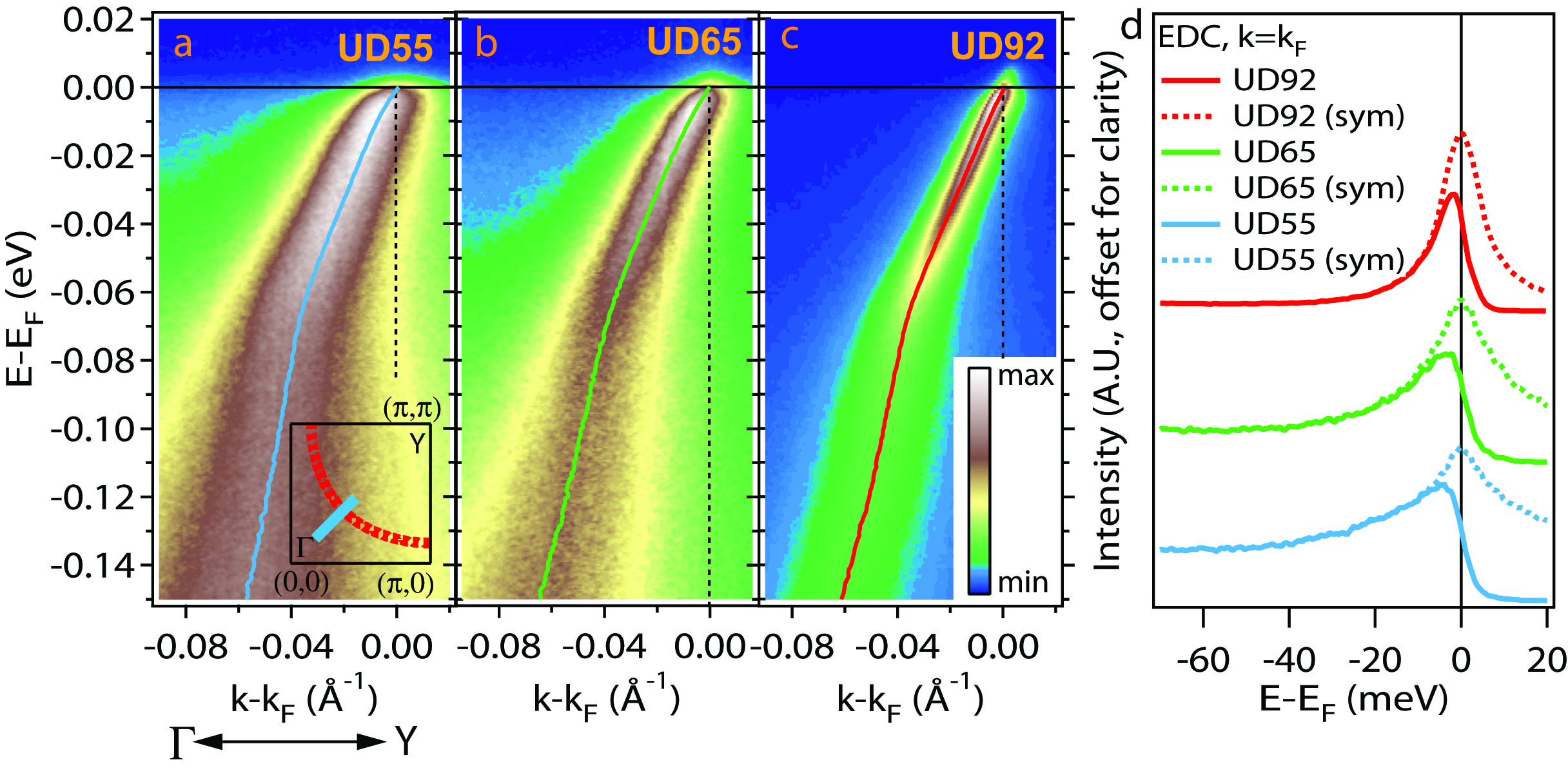}
\centering
\caption{\label{Fig 1: Nodal dispersion in Bi-2212}  (a)-(c) False-color image plots of the nodal dispersion of UD55, UD65, and UD92, measured at 10K.  Solid curves indicate band dispersions derived from MDC peak positions.  Inset of (a): Brillouin zone schematic with Fermi surface indicated in red.  The \textit{x}-axes in (a)-(c) correspond to momentum along diagonal blue line (`nodal cut') in inset. (d) EDCs (solid) and symmetrized EDCs (dashed) at k$_F$. A single peak at E$_F$ in the latter confirms that spectra in (a)-(c) are ungapped. k$_F$ determined from Fermi crossing of dispersion (vertical dashed lines in (a)-(c)).}
\end{figure}

In a \textit{d}-wave superconductor like the high-T$_c$ cuprates, the electronic component of low-temperature thermodynamics is dictated by the nodes, where arbitrarily small excitations are permitted by the gapless nature of these points.  An intriguing aspect of cuprate phenomenology is the so-called universal nodal Fermi velocity (v$_F$).\cite{UniversalNodalvF}  Along the nodal direction ((0,0)-($\pi$,$\pi$)) the velocity measured by ARPES within 50 meV of E$_F$ appears to be independent of cuprate-family or the number of CuO$_2$ layers in the compound, and is also nearly constant across the phase diagram -- from the undoped insulator, across the superconducting dome, and in the non-superconducting metallic state at a doping \textit{p}$>$0.25 -- even though other electronic properties vary significantly with doping.\cite{PhaseDiagramInPlaneResistivity:Ando,DopingDependencePG:Tallon,DopingDependenceLSCO:Yoshida} In addition, this universal v$_F$, if combined with ARPES measurement of the superconducting gap, leads to apparent contradiction with thermal conductivity observed directly in transport measurements,\cite{InhomogeneityBreakdownThermalConductivity:Sun, TheoryBreakdownUniversalThermal:AndersenHirchfeld} suggesting that crucial information about the nodal quasiparticles is still missing.

ARPES data can be represented as a convolution between the single-particle spectral function and the momentum and energy resolution of the experiment. Naturally, with the improved resolution of laser-based ARPES, the measured spectrum begins to approach the intrinsic spectral function, and finer structure can be revealed.  Recent laser-ARPES measurements along the nodal direction of optimally-doped Bi-2212 have uncovered a low-energy ($<$10 meV) kink,\cite{LaserARPES:LEKinkPlumb} in addition to the larger kink seen at 50-80 meV in all cuprates.\cite{ElectronPhononCoupling:Lanzara,KinkPRL:Bogdanov,Zhou:EPhCoupling} Other laser-ARPES studies have shown a corresponding decrease in nodal linewidth at low energies.\cite{LaserARPES:NodalQPIshizaka,Bi2212NewCouplingLaser:Zhang,Rameau:LEkink}  In this letter, we present the systematics of the low-energy kink by means of a doping-dependent study of underdoped Bi-2212.  20 samples with 6 dopings in the range 0.076$<$\textit{p}$<$0.14 were measured using a 7 eV  laser and a Scienta SES2002 analyzer.  7 eV photons were produced by second harmonic generation from a 355 nm laser (Paladin, Coherent, Inc.) using a nonlinear crystal KBe$_2$BO$_3$F$_2$.\cite{KBBF:Liu} Energy and momentum resolution were 3 meV and  better than 0.005 \AA$^{-1}$, respectively. Samples were cleaved in-situ at a pressure $<$4$\times$10$^{-11}$ torr to obtain a clean surface, and measured at 10K.

\begin{figure} [t]
\includegraphics [type=jpg,ext=.jpg,read=.jpg,clip, width=3.3 in]{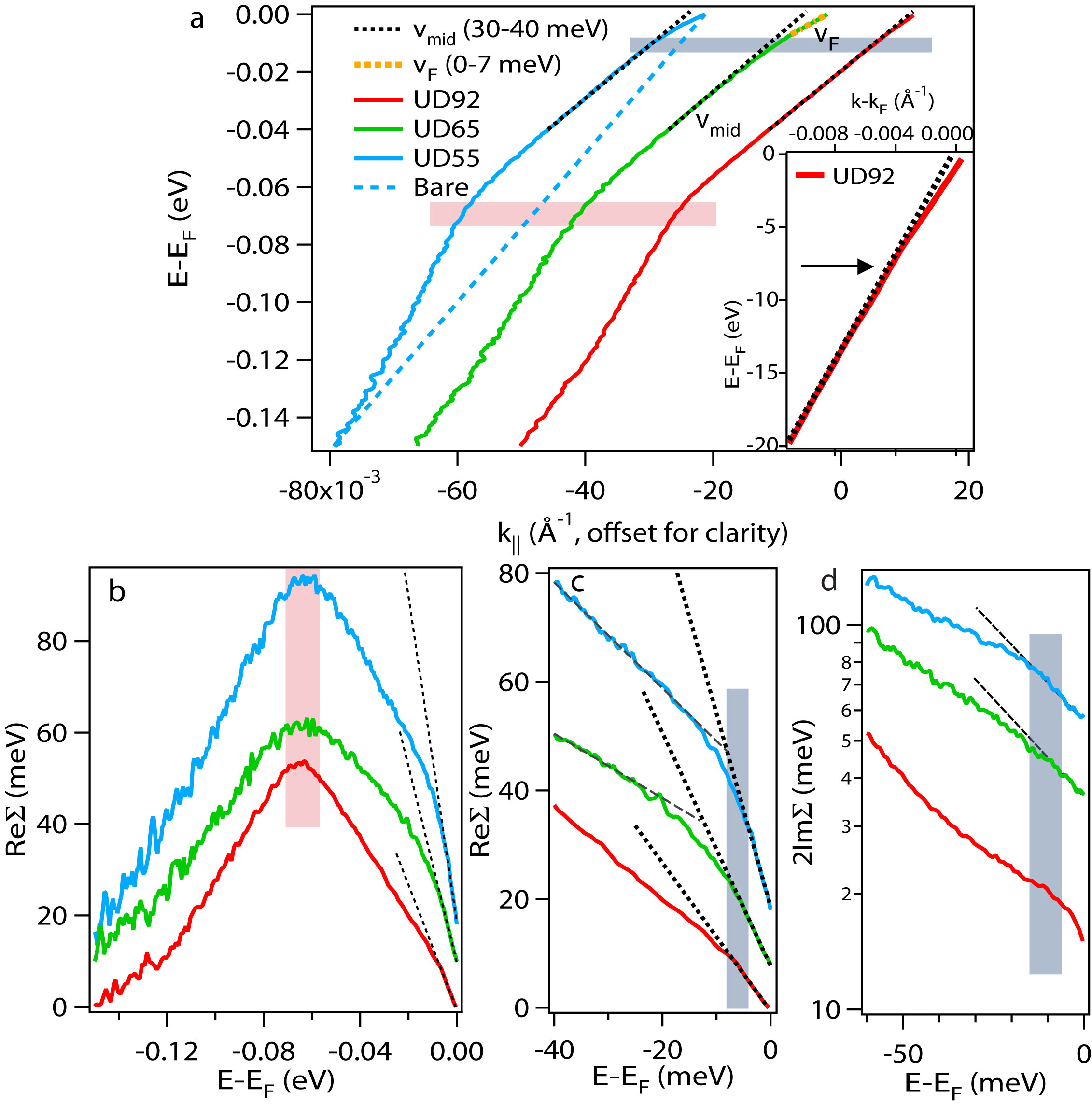}
\centering
\caption{\label{Fig 2: Systematics of low-energy kink} (a) Band dispersions from Fig \ref{Fig 1: Nodal dispersion in Bi-2212}, offset for clarity.  Dashed line accompanying UD55 denotes assumed linear bare band. Black dotted lines are linear fits 30-40 meV (v$_{mid}$), extrapolated to E$_F$. This differs from v$_F$ (fit 0-7 meV), indicated on UD65 dispersion by orange dashed line. Pink bar marks 70 meV kink, and grey bar marks low-energy kink, where v$_F$ deviates from v$_{mid}$. Inset: UD92 dispersion below 20 meV. (b) Re$\Sigma$, approximated by subtracting a linear bare band from dispersions in (a), is peaked at the position of the 70 meV kink and changes slope near the energy position of the low-energy kink.  (c) Detail of the low-energy portion of Re$\Sigma$.  Thick dotted lines are fits near E$_F$.  All curves deviate from these lines between 6-10 meV, as highlighted by the shaded bar.  For the two most underdoped samples, the slope of Re$\Sigma$ evolves between the low-energy kink and 20-30 meV, suggesting an additional kink. (d) 2Im$\Sigma$, with dotted lines as guides-to-the-eye.  All curves decrease more rapidly near E$_F$, but most markedly in UD92.  For more underdoped samples, the effect may be obscured by a larger linewidth and possible additional kink.}
\end{figure}

Fig. \ref{Fig 1: Nodal dispersion in Bi-2212}(a)-(c) show ARPES image plots for nodal cuts at three dopings: UD55 (underdoped, T$_c$=55K), UD65, and UD92, corresponding to hole-dopings of approximately  0.088, 0.10, and 0.14. Standard momentum distribution curve (MDC) analysis--Lorentzian fits at fixed energy-- is used to extract the band dispersions.\cite{SelfEnergyEffectsPhotoemission:Johnson}  The energy distribution curves (EDCs)--intensity as a function of energy at fixed momentum-- at k$_F$ in Fig. \ref{Fig 1: Nodal dispersion in Bi-2212}(d) indicate that these spectra are ungapped, as the symmetrized EDCs \cite{Symmetrization_Norman_model} have a single peak at E$_F$.


The systematics of the low-energy kink are studied via the MDC-derived nodal dispersion, which are plotted for three dopings in Fig. \ref{Fig 2: Systematics of low-energy kink}(a).  In addition to the large, ubiquitous kink near 70 meV, a smaller kink is also evident: the dispersion within 10 meV of E$_F$ deviates from the velocity fit between 30-40 meV.  This deviation appears more pronounced for more underdoped samples.  Consistent with the work of Plumb \textit{et al}, the velocity (slope of the MDC dispersion) within 7 meV of E$_F$ is \textit{smaller} than the velocity at higher binding energy, notably opposite to the expected effects of instrument  and thermal broadening.\cite{LaserARPES:LEKinkPlumb}  We also note that the low-energy kink cannot be identified as an artifact due to a gap, because measurements are performed at the node where the superconducting gap is zero.

Another way to visualize the low-energy kink is via the real part of the electronic self-energy, Re$\Sigma$, plotted in Fig. \ref{Fig 2: Systematics of low-energy kink}(b)-(c).  The low-energy kink is marked by a deviation of the slope of Re$\Sigma$ at 6-10 meV from the slope established at E$_F$.  For the sample closest to optimal doping, UD92, there is a single `knee' in Re$\Sigma$.  Meanwhile, the slope of Re$\Sigma$ for UD55 and UD65 continues to evolve until 20-30 meV, possibly suggesting an additional kink, reminiscent of the 70meV kink, which may have several components.\cite{MultipleKinks:Lee,AspectsEPhCoupling:Lee,MultipleModes:XJ}  From the Kramers-Kronig relation between Re$\Sigma$ and Im$\Sigma$, a signature of the low-energy kink is expected in Im$\Sigma$, which is proportional to the MDC FWHM.  In Fig. \ref{Fig 2: Systematics of low-energy kink}(d) we show that all dopings exhibit a downturn in Im$\Sigma$ near E$_F$, though this is most pronounced for UD92.  For more underdoped samples, the larger linewidth and possible additional kink make it more difficult to get quantitative information from Im$\Sigma$, but the observation that Im$\Sigma$ decreases more rapidly close to E$_F$ remains robust.   The appearance of a low-energy feature in both Re$\Sigma$ and Im$\Sigma$ strongly argues against a spurious origin for the low-energy kink, and  the phenomenology reported in Fig. \ref{Fig 2: Systematics of low-energy kink} is reproduced in the other samples in our study.  Thus, the systematics of a new energy scale can be added to the hierarchy of multiple energy scales in the cuprates.\cite{HierarchyManyBodyInteractions:Meevasana}



\begin{figure} [t]
\includegraphics [type=jpg,ext=.jpg,read=.jpg,clip, width=3.3 in]{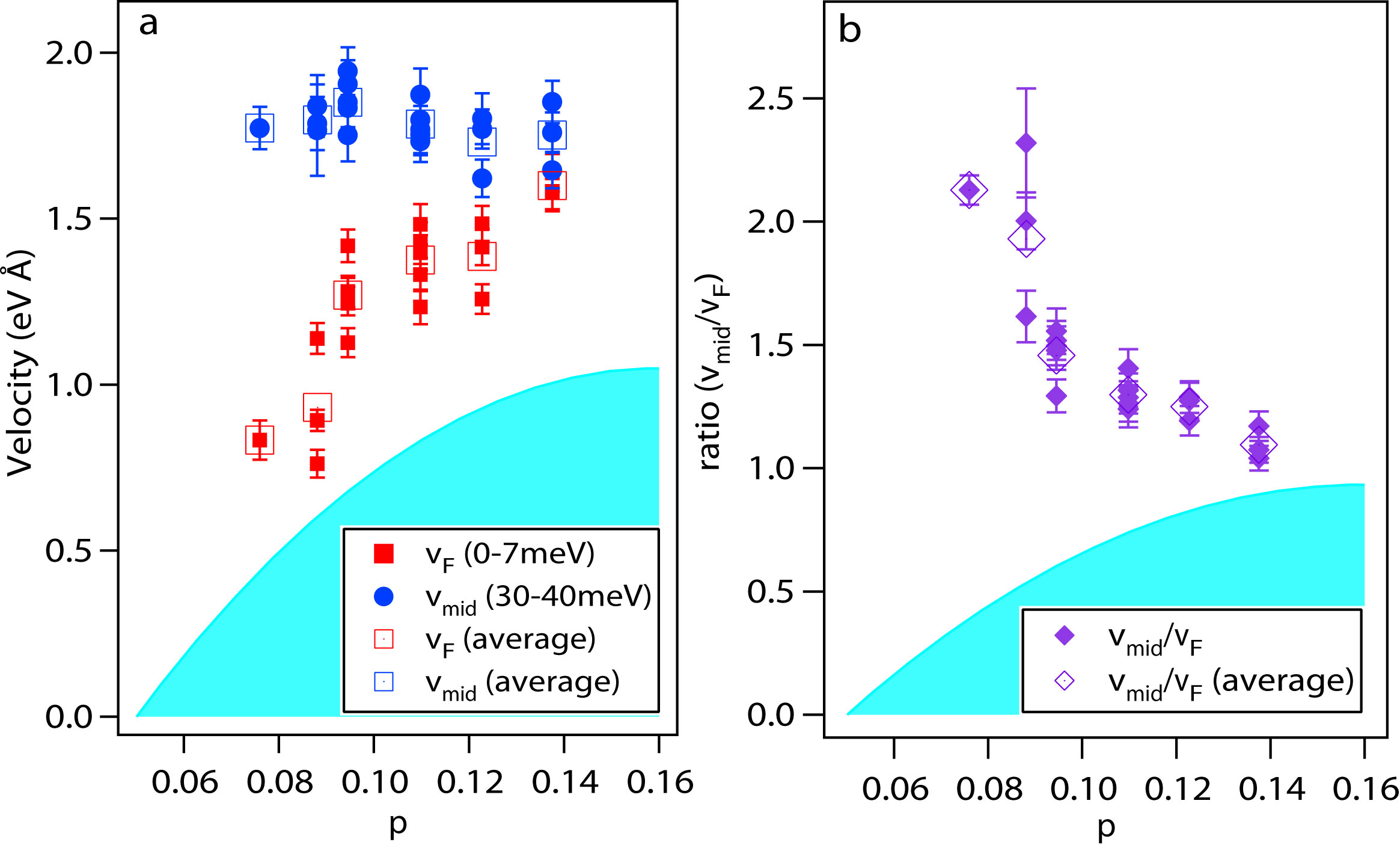}
\centering
\caption{\label{Fig 3: Non-universal nodal Fermi velocity} (a) Doping dependence of v$_F$ and v$_{mid}$. Open red (blue) squares denote average v$_F$ (v$_{mid}$) for each doping.  Boundary of shaded region denotes doping dependence of T$_c$. Doping is approximated from empirical relation T$_c$$=$96(1-82.6(p-0.16)$^2$).\cite{UniversalCurve} v$_{mid}$ has little systematic doping dependence, while v$_F$ decreases with underdoping.  (b) Doping dependence of v$_{mid}$/v$_F$, which is related to the renormalization coupling strength. }
\end{figure}

The ubiquity of the low-energy kink in UD Bi-2212 leads us to reexamine previous measurements of v$_F$, as there is now compelling evidence that quasiparticles very close to E$_F$ experience a heretofore unconsidered mass renormalization.  The  nodal v$_F$ is plotted in Fig. \ref{Fig 3: Non-universal nodal Fermi velocity}, and our key finding is that v$_F$ is not universal, but rather, has a pronounced doping dependence in the regime of this study. To characterize our data, at least two velocities are needed: v$_{mid}$, the linear fit between 30-40 meV, and v$_F$, the velocity fit between 0-7 meV, as defined in Fig. \ref{Fig 2: Systematics of low-energy kink}(a). These energy ranges are chosen to get sufficient data points while avoiding the low-energy kink and 70 meV kink.  v$_{mid}$ is found to be approximately 1.8 eV\AA, without a distinct doping dependence,  consistent with the previously reported `universal' value.\cite{UniversalNodalvF}  Meanwhile, \textit{v$_F$ decreases monotonically with underdoping}.  This is consistent with recent quantum oscillation results, which suggest a divergence of the effective mass in the underdoped regime.\cite{Sebastian:DivergentMass} The coupling strength of this low-energy renormalization can be roughly assessed by the velocity ratio v$_{mid}$/v$_F$, which is plotted in Fig. \ref{Fig 3: Non-universal nodal Fermi velocity}(b), and suggests that coupling strength increases with underdoping.  Notably, the ratio of velocities on either side of the 70meV kink exhibits the same doping dependence, though with the 70 meV kink, it is the higher energy velocity ($\omega$$>$70meV) which is doping-dependent.\cite{UniversalNodalvF}  Although a doping-dependent v$_F$ presents a significant shift from our previous understanding of cuprate nodal physics, our results are not inconsistent with previous measurements: v$_{mid}$ is indeed doping-independent, and inferior energy resolution can easily obscure subtle low-energy kinks near E$_F$.  Our finding underscores the importance of very low energy scales in these systems and revises cuprate phenomenology by linking nodal v$_F$ to doping and T$_c$, previously suggested by the temperature dependence of the low-energy kink.\cite{LaserARPES:LEKinkPlumb} Further, this doping dependence constrains the origin of the low-energy kink, and may aid interpretation of bulk thermodynamic measurements, particularly thermal conductivity, which will be the focus of the remaining discussion.

\begin{figure*} [t]
\includegraphics [type=jpg,ext=.jpg,read=.jpg,clip, width=6.5 in]{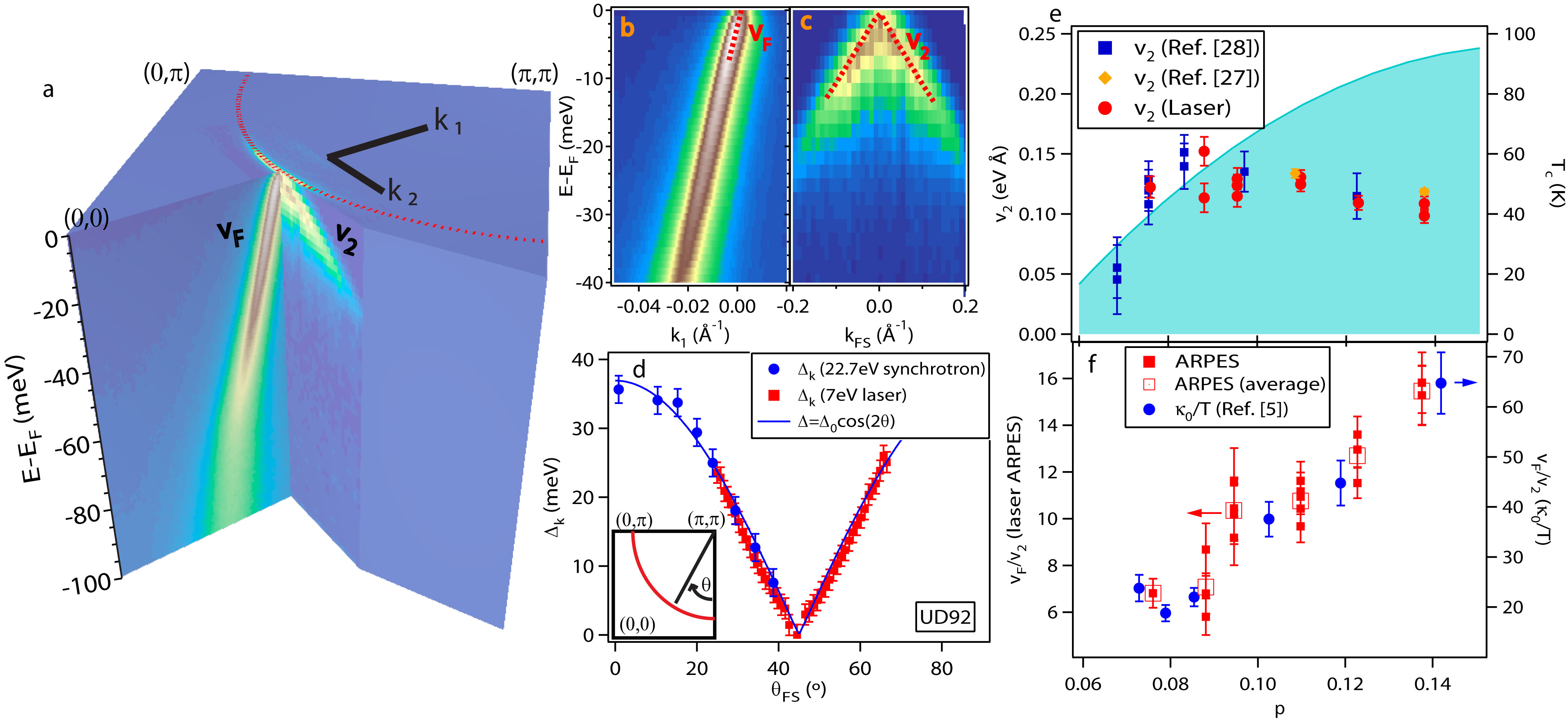}
\centering
\caption{\label{Fig 4: Low energy excitations and comparison to thermal conductivity} (a) Cutout showing FS (top) and dispersion perpendicular (v$_F$, left) and tangential (v$_2$, right) to FS at node, from measurement on UD92 at 10K. (b)-(c) Image plots showing measured v$_F$ and v$_2$ directly.  The latter image consists of EDCs at k$_F$ from many parallel cuts near the node.  (d)  Comparison of synchrotron- and laser-based ARPES measurements of the superconducting gap of UD92 around the FS. When the data in (d) is fit to a simple d-wave form, $\Delta$($\theta$)=$\Delta$$_0$$\cos$(2$\theta$) close to the node,  v$_2$ $\approx$ 2$\Delta$$_0$/k$_F$, where k$_F$ is the distance from the node to ($\pi$,$\pi$). (e) v$_2$ from synchrotron- and laser-based ARPES experiments. (f) Comparison between v$_F$$/$v$_2$ from laser ARPES and thermal conductivity from Ref. \cite{InhomogeneityBreakdownThermalConductivity:Sun}.  Doping dependence is consistent, though absolute values differ. Open squares indicate average laser ARPES values.}
\end{figure*}

 For the cuprates, thermal conductivity near T=0 can be expressed in terms of two components of the Fermi velocity: the velocity perpendicular to (v$_F$) and tangential to (v$_2$) the Fermi surface (FS) at the node (Fig. \ref{Fig 4: Low energy excitations and comparison to thermal conductivity}(a)-(c)).\cite{LowEnergyQP:Chiao,LowEnergyQPtheory:Hussey,HeatTransportProbeGapStructure:Shakeripour}  For a 2D \textit{d}-wave superconductor in the clean limit, the residual linear term (T=0 extrapolation) of  thermal conductivity, $\kappa$$_0$/T, is independent of the quasiparticle scattering rate, interaction energy, or other sample-dependent parameters.\cite{LowEnergyQPtheory:Hussey,QPTransport:DurstLee}

In this regime, $\kappa$$_0$/T is related to v$_F$ and v$_2$ by a simple formula:\cite{LowEnergyQPtheory:Hussey,QPTransport:DurstLee,LowEnergyQP:Chiao}
\begin{equation}\label{1}
\kappa_0/T=\frac{k_B^2}{3\hbar}\frac{n}{d}[\frac{v_F}{v_2}+\frac{v_2}{v_F}],
\end{equation}
where \textit{n} is the number of CuO$_2$ planes per unit cell, and \textit{d} is the c-axis unit cell length.  The second term is usually negligible, as v$_2$$\ll$v$_F$. Thus, by measuring bulk thermal conductivity, one can extract a microscopic parameter, v$_2$$/$v$_F$, which fully determines the ground state nodal electronic structure of cuprates.\cite{HeatTransportProbeGapStructure:Shakeripour} Thermal conductivity measurements on La$_{2-x}$Sr$_x$CuO$_4$,\cite{LSCOThermalTransport:Takeya,ThermalConductivityPhaseDiagram:Sutherland} YBa$_2$Cu$_3$O$_7$,\cite{ThermalConductivityPhaseDiagram:Sutherland} Bi$_2$Sr$_{2-x}$La$_x$CuO$_{6+\delta}$,\cite{QPTBi2201:Ando} and Bi-2212\cite{InhomogeneityBreakdownThermalConductivity:Sun} have all shown that  $\kappa$$_0$/T decreases with underdoping, implying that v$_F$$/$v$_2$ also decreases. However, using v$_2$ reported in Ref. \cite{SCGapBi2212:Mesot} and a universal v$_F$, Sun \textit{et al} argued that ARPES suggested a different doping dependence of v$_F$/v$_2$. \cite{InhomogeneityBreakdownThermalConductivity:Sun} Such contradictions have been attributed to disorder effects, such as electronic inhomogeneity \cite{InhomogeneityBreakdownThermalConductivity:Sun} or disorder-induced magnetism,\cite{TheoryBreakdownUniversalThermal:AndersenHirchfeld} but previous analysis lacked a crucial component: a doping-dependent v$_F$.

For comparisons via Eqn. (1), we have obtained v$_2$ from laser-ARPES measurements of the momentum dependence of the superconducting gap near the node (Fig. \ref{Fig 4: Low energy excitations and comparison to thermal conductivity}(d)), and these values are consistent with recently published data. \cite{ARPES:Tanaka,ARPES:twoGap_WS}  Using these v$_2$ together with v$_F$ from Fig. \ref{Fig 3: Non-universal nodal Fermi velocity}(a),  we plot the ratio v$_F$$/$v$_2$  in Fig \ref{Fig 4: Low energy excitations and comparison to thermal conductivity}(f) alongside the thermal conductivity values reported by Sun \textit{et al}.\cite{InhomogeneityBreakdownThermalConductivity:Sun}  The ARPES v$_F$$/$v$_2$ decrease strongly with underdoping, exhibiting a consistent trend with the thermal conductivity results for Bi-2212 as well as other cuprates.  The v$_F$$/$v$_2$ derived from ARPES and thermal conductivity differ in absolute value, and this may be related to the in-plane anisotropy of thermal conductivity that has been reported in Bi-2212 ($\kappa$$_a$$\neq$$\kappa$$_b$). \cite{InhomogeneityBreakdownThermalConductivity:Sun,AnisotropyThermalConductivityBi2212:Ando} The doping-dependent data in Ref. \cite{InhomogeneityBreakdownThermalConductivity:Sun} were along the \textit{a}-axis, which has a larger $\kappa$$_0$/T, but the discrepancy in absolute value of v$_F$/v$_2$ shown in Fig. \ref{Fig 4: Low energy excitations and comparison to thermal conductivity} (f) may indicate that the \textit{b}-axis may be the correct one to compare to ARPES.  Notably ARPES does not observe in-plane anisotropy of nodal single particle parameters (v$_F$ and v$_2$), so the anisotropy in thermal conductivity must have a different origin, such as coexisting density-wave order.\cite{Theory:DensityWave} Alternately, the difference in absolute value of v$_F$/v$_2$ may suggest a proportionality constant of different origin is needed in Eqn. (1). Nevertheless, with the doping-dependent v$_F$ elucidated by laser ARPES, we are able to reproduce the doping-dependence of $\kappa$$_0$/T which is seen in multiple cuprate families.

The superior resolution of laser ARPES allows access to some of the lowest energy scales of high-T$_c$ cuprates, and the findings may further constrain the microscopic theory of high-T$_c$ superconductivity.  With this powerful probe, we have been able to uncover the doping dependence of a microscopic parameter, v$_F$.  A non-universal nodal Fermi velocity both answers old questions and introduces new ones: discrepancies between doping-dependent ARPES and thermal conductivity measurements can be resolved, but there are new questions about the origin of the low-energy kink, the explanation of its doping and temperature dependence, and its role in superconductivity.

We thank Profs. N. Nagaosa,  J. Zaanen, and F. Ronning for helpful discussions.  SSRL is
operated by the DOE Office of Basic Energy Science. This work is supported by DOE Office of Basic Energy Science,
Division of Materials Science, with contracts DE-FG03-01ER45929-A001 and DE-AC02-76SF00515.

\bibliography{LE_kink}
\end{document}